
\def\ifundefined#1{\expandafter\ifx\csname
#1\endcsname\relax}

\newcount\eqnumber \eqnumber=0
\def\beq{ \global\advance\eqnumber by 1 $$ }
\def\eeq{ \eqno(\the\eqnumber)$$ }
\def\label#1{\ifundefined{#1}
\expandafter\xdef\csname #1\endcsname{\the\eqnumber}
\else\message{label #1 already in use}\fi}
\def\(#1){(\csname #1\endcsname)}
\def\puteqno{\global\advance \eqnumber by 1 (\the\eqnumber)}

\def\label#1{
\ifundefined{#1}
\expandafter\edef\csname #1\endcsname{\the\eqnumber}
\else\message{label #1 already in use}
\fi{}}
\def\(#1){(\csname #1\endcsname)}
\def\eqn#1{(\csname #1\endcsname)}

\newcount\refno \refno=0
\def\[#1]{\ifundefined{#1}\advance\refno by 1
\expandafter\xdef\csname #1\endcsname{\the\refno}
\fi[\csname #1\endcsname]}
\def\refis[#1]{\item{\csname #1\endcsname.}}

\def\sqr#1#2{{\vcenter{\vbox{\hrule height.#2pt
        \hbox{\vrule width.#2pt height#1pt \kern#1pt
          \vrule width.#2pt}
        \hrule height.#2pt}}}}
\def\square{\mathchoice\sqr68\sqr68\sqr{4.2}6\sqr{2.10}6}

\def\n{\noindent}


\baselineskip=18pt
\magnification=1200

\def\d#1/d#2{ {\partial #1\over\partial #2} }

\newcount\sectno

\newcount\subsectno

\def\subsect{\global\advance\subsectno by1 \the\sectno.\the\subsectno }
 \def\sect{\subsectno=0 \global\advance\sectno by1 \the\sectno }

\def\pdr{\partial}

\def\eps{\epsilon}
\def\half{{1\over 2}}

\def\Tr{\hbox{Tr}}

\def\linebreak{\hfil\break}


\newcount\eqnumber
\def\beq{ \global\advance\eqnumber by 1 $$ }
\def\eeq{ \eqno(\the\eqnumber)$$ }
\def\puteqno{
\global\advance \eqnumber by 1 (\the\eqnumber)}
\def\beqs{$$\eqalign}
\def\eeqs{$$}


\def\ifundefined#1{\expandafter\ifx\csname
#1\endcsname\relax}

\newcount\refno \refno=0  
\def\[#1]{
\ifundefined{#1}
\advance\refno by 1
\expandafter\edef\csname #1\endcsname{\the\refno}
\fi[\csname #1\endcsname]}
\def\refis#1{\noindent\csname #1\endcsname. }

\def\label#1{
\ifundefined{#1}
\expandafter\edef\csname #1\endcsname{\the\eqnumber}
\else\message{label #1 already in use}
\fi{}}
\def\(#1){(\csname #1\endcsname)}
\def\eqn#1{(\csname #1\endcsname)}

\baselineskip=15pt
\parskip=10pt


 \def\diff{\;{\rm Diff} \;}
\def\semi{\otimes}
\def\Q{{\cal Q}}
\def\Lie{{\cal L}}

\def\un#1{\underline{#1}}

\def\diff{\;{\rm Diff} \;}
\def\semi{\otimes}
\def\Q{{\cal Q}}
\def\Lie{{\cal L}}

\def\un#1{\underline{#1}}


						\hfill June 2, 1994

			\hfill Revised Sept. 27, 1994; Nov. 18, 1994

						\hfill UR-1357

						\hfill ER-40685-807

\vskip0.1in
\centerline
{\bf O(N) Sigma Model as a Three Dimensional Conformal Field Theory}

\vskip 0.1in
\centerline {\dag S.~Guruswamy, \dag S.~G.~Rajeev
and \dag\dag P.~Vitale}
\vskip0.2in
\centerline
{\dag Department of Physics, University of Rochester}
\centerline
 {Rochester, NY 14627 USA}
\centerline {e-mail: guruswamy, rajeev@urhep.pas.rochester.edu}
\centerline {and}
\vskip0.1in
\centerline
{\dag \dag Dipartimento di Scienze Fisiche, Universit\`a di Napoli}
\centerline
{and I.N.F.N. Sez.\ di Napoli,}
\centerline
{Mostra d'Oltremare Pad.~19, 80125 Napoli ITALY.}
\centerline{e-mail: vitale@axpna1.na.infn.it}
\vskip0.2in
\centerline{\bf Abstract}

We study a three dimensional conformal field theory
in terms of its partition function on arbitrary
curved spaces. The large $N$ limit of the nonlinear sigma model at the
non-trivial fixed point is shown to be an example of a conformal
field theory, using zeta--function regularization.
We compute the critical properties of this model in various
spaces of constant curvature ($R^2 \times S^1$, $S^1\times S^1 \times R$,
 $S^2\times R$,
$H^2\times R$, $S^1 \times S^1 \times S^1$
and $S^2 \times S^1$) and we argue that what distinguishes the
different cases is not the Riemann curvature but the conformal
class of the metric.  In the case $H^2\times R$
(constant negative curvature),
the $O(N)$ symmetry is
spontaneously broken at the critical point.
 In the case
$S^2\times R$ (constant positive curvature) we find that the
free energy vanishes,
consistent with conformal equivalence of this manifold to $R^3$,
although the correlation length is finite. In the zero curvature cases,
the correlation length is finite due to finite size effects.
These results describe two dimensional quantum phase transitions or
three dimensional classical ones.

\vfill
\eject

\noindent {\bf \sect. Introduction}

The correlation functions of  statistical mechanical systems
exhibit scale invariance at a phase transition of second or
higher order. It is possible to describe such a  phase
transition in terms of a scale invariant euclidean quantum
field theory, a typical   classical statistical system
corresponding to a regularized euclidean quantum field theory.
The correlation length (inverse  mass of the particles) of
this field theory will diverge as the coupling constants
approach a `fixed point'. The theory defined by the
fixed point in  the space of  coupling constants will be scale
invariant and corresponds to a phase transition point of the
statistical mechanical system. The immediate neighborhood of
the fixed point describes   quantum field theories with masses of the
particles
small compared to the  cut--off.

At present, the only general  way to construct  quantum field
theories is as such limiting cases of statistical mechanical
systems. It is of great interest to construct   scale
invariant quantum
field theories directly. This should be possible since they
are `finite' (i.e., correspond to  a fixed point of the
renormalization group); moreover it could eventually lead to a
definition of a general quantum field theory
as a perturbation to a scale invariant quantum field theory
without the usual cumbersome procedure of renormalization.
Any progress in that direction is of  importance to particle physics, where the
standard model  should eventually be constructed by such an
intrinsic procedure.

Scale invariant
quantum field theories are also  interesting for
phenomenological reasons:  they describe
experimentally accessible phase transition phenomena.
The critical exponents of such transitions have been
calculated with great accuracy in many realistic cases.
Also,
in the case of two dimensional systems,  many systems have
been solved exactly. That is, their partition and correlation
functions have been obtained in terms of the special functions
of classical mathematics.

The key to this success in two dimensions is that the theories
often have a symmetry much larger than scale invariance:
conformal
invariance. In two dimensions, infinitesimal conformal
transformations (position dependent scale transformations)
are
the same as complex analytic co--ordinate transformations.
This large symmetry puts strong  constraints on the
correlation
functions; in the case of `minimal models' these constraints
are strong enough to determine them completely. This is
similar in spirit
to the solution of classical potential problems in two
dimensions by conformal (Schwarz--Christoffel)
transformations. For example, the general boundary value
problem of the
two dimensional Laplace equation can be solved by mapping the
boundary to one of a small class of standard boundaries.

Most interesting classical phase transitions occur in three
dimensional systems. One should not expect a complete
generalization  of conformal techniques to three dimensions
(as an analogy, it is not possible to solve the general
boundary value problem of the Laplace equation in three
dimension by conformal transformation to a standard boundary).
One should expect that conformal invariance is a powerful
constraint even in this case, although not strong enough to
completely determine the correlation functions. In this paper
we will begin a study of three dimensional conformal field
theory. (We study the $O(N)$ non-linear sigma model).
It is seen that in three dimensions also, the
partition and correlation functions at a second or higher
order transition are invariant under conformal
transformations of the metric tensor.
We will also compute the
partition function in some special cases  to confirm this
general
picture.

In three dimensions, it is necessary
to study field theories in a curved space to fully understand
conformal invariance. In two dimensions all manifolds are conformally
flat, so the only geometric information about the manifold
that can affect critical systems is a finite number of
(Teichmuller)  moduli parameters.
This is replaced in the three dimensional case by an
infinite dimensional space of conformal structures
(see appendix A for a precise definition). The partition function
of the system should be viewed as a functional of the
metric density $\hat g_{ij}=g^{-{1 \over d}} g_{ij}$, where
 $g={\rm det}g_{ij}$; it is the generating functional
of the correlation functions of the stress tensor.
The point is that conformal invariance alone will not
determine this functional, even for `minimal' models.
The correlation functions of the stress tensor are physically
measurable quantities; therefore  the partition function on
curved spaces is of physical interest even
when the true metric of space is flat.

There is another reason to study critical phenomena
in curved spaces. If we subject a system to external stresses,
we can deform the underlying microscopic structure
(such as a lattice) so that the effective distance between
points is no longer the usual one.  At  a critical point, universality
suggests that the details of the microscopic structures
do not matter; but the system is still sensitive to the deformation
through the effective metric tensor density.

In three dimensions there is no conformal anomaly;
hence a conformal field theory can be defined as
one whose partition function is a function
on the space of conformal structures
(if there were an anomaly, it would  have been a
section of a real line bundle).
That is, its partition function should satisfy
\beq
	{\cal Z}[e^{2f}g]={\cal Z}[g]
\eeq
$g$ being the metric tensor.
It is not obvious that interacting  theories of
this kind exist; we will show that the large
$N$ limit of the $O(N)$ sigma model \[arefeva],\[polyabook],\[sachdev],
\[rosenstein],\[zinnjustin]
is such a
conformal field theory, at its non-trivial fixed point.
Moreover a primary field is one
whose correlation functions transform homogeneously under
a conformal transformation:
\beq
	<\phi(x_1)\cdots \phi(x_n)>_{e^{2f}g}
=e^{\Delta(f(x_1)+\cdots f(x_n))}	<\phi(x_1)\cdots \phi(x_n)>_{g}
\eeq
Again, the field $\phi_i$ of the $O(N)$ sigma model
is an example of such a primary field.
These  definitions are the appropriate
generalizations of the point of view of Ref.\[jain]. For previous work on
 conformal field theory in dimensions greater than two, see Refs.
\[witten1],\[osborn],\[latorre],\[mottola].

Although there is no conformal anomaly  in three
dimensions, there could be  a parity anomaly in general.
The the particular example we are studying, the $O(N)$ sigma model
does not have such an anomaly.
We plan to return to this issue in the context of
fermionic and Grassmannian sigma  models \[ferretti].

The following is the lay-out of this paper. Sect. 2 is a discussion of the
large $N$ limit of the
$O(N)$ sigma model on arbitrary three-dimensional manifolds. Sect. 3 is
a brief review of the results on $R^3$. Sect. 4 discusses the divergence
structure of the theory. Sect. 5 describes the zeta function
regularization and an outline of the proof of the conformal invariance
of the free energy density is given. Sect. 6 and Sect. 7 contain the study of
the theory on special manifolds.
Appendix A is a pedagogic discussion of conformal
geometry in three dimensions. This appendix contains material for the
reader who is interested in the geometric
context for the definition of a conformal field theory in three dimensions. In
particular the differences between two and three dimensional conformal geometry
 and the special nature of conformal curvature in three dimensions are
stressed.
Appendix B is a derivation of the Poisson sum
formula on
$S^2$. A table summarizing the results is given at the end of the last
section.

\n {\bf \sect. O(N) non-linear sigma model in three dimensions}

In this section we study the $O(N)$ non-linear sigma model [1-5]
on a Riemannian manifold $(M,g)$.
Of particular interest to us is the case
$M=\Sigma \times R$ with $\Sigma$ being an arbitrary 2--dimensional curved
space: this describes a
quantum phase transition at zero temperature, the temperature being the
inverse radius of $R$.
We will study the theory in detail at its non-trivial fixed point for
manifolds
$\Sigma$ with zero, constant positive and constant negative Riemannian
curvature.

The euclidean partition function of the $O(N)$ non-linear
sigma model in
3--dimensions in the presence of a background metric $g^{\mu \nu}(x)$is
given by,
\beq {\cal Z}[g]=\int {\cal D[\phi]} e^{{-1\over 2\lambda} \int {d^3x
{\surd g(x)}[g^{\mu \nu}(x) \partial_{\mu} \phi^i \partial_{\nu}\phi_i]}}\eeq
where $i=1,2,\cdots,N$ and the $\phi^i(x)$ satisfy
the constraint $\phi^i(x) \phi_i(x)=1$. $\lambda$ is a coupling constant.
As it stands, the action is not invariant under the conformal transformation
of the metric,
$g^{\mu \nu}(x)\rightarrow \Omega^{2}(x) g^{\mu \nu}(x)$.
The action becomes conformally invariant \[birrell] when the laplacian
$-{{1\over{\surd g}}\partial_{\mu}({\surd g(x)}g^{\mu\nu}(x)\partial_{\nu})}$
is modified to the ``conformal laplacian'',
$-\square_g=
{-{1\over{\surd g}}\partial_{\mu}({\surd g(x)}g^{\mu\nu}(x)\partial_{\nu})}
+\xi {\rm R}$, with $\phi$
transforming as
$\phi(x)\rightarrow \Omega^{1-{d\over 2}}(x) \phi(x)$.
${\rm R}$, in the conformal laplacian, denotes the
Ricci scalar and $\xi$ is a numerical constant.
$\xi={{d-2}\over {4 (d-1)}}$ and is equal to $1/8$ for dimensions $d=3$.
The generating functional then reads,
$${\cal Z}[g]=\int {\cal D[\phi]} e^{{-1\over 2\lambda} \int {d^3x
{\surd g}[-\phi ^i \square_g \phi_i]}}$$
However, the constraint $\phi^i(x) \phi_i(x)=1$
still violates conformal invariance.
The constraint on the $\phi$ fields can be implemented by a
Lagrange multiplier, in the form of an auxiliary field $\sigma(x)$ with no
dynamics, as follows:
\beq {\cal Z}[g]=\int {\cal D[\phi] D[\sigma]} e^{{-1\over 2\lambda} \int {d^3x
 {\surd g}[-\phi^i\square_g {\phi_i}+{\sigma}({\phi^i}^2-1)]}}.\eeq
Although, classically this is not conformally invariant,
we will show that there
is a non-trivial fixed point for the quantum theory at which it is conformally
invariant.
Following Wilson's approach, we regularize the generating functional $\cal Z$
in the ultraviolet by introducing a cut-off, $\Lambda$, in the momentum
space. Before doing that, let us note the canonical dimensions of the fields
and the couplings in our action. Rescaling $\phi^i$ to
${\surd \lambda} \phi^i$, we have,
$${\cal Z}[g]=\int {\cal D[\phi] D[\sigma]} e^{- \int {d^3x
{\surd g}[{-1\over 2} {\phi^i}\square_g {\phi_i}
+{1\over 2}{\sigma}{\phi^i}^2-{{\sigma}\over 2\lambda}]}}.$$
The canonical dimensions in mass units can be read off from the action:
$$[\phi]=1/2\quad, \quad [\sigma]=2 \quad,\quad [{1\over \lambda}]=1.$$
Let us redefine our coupling constant $1\over \lambda(\Lambda)$ to
${\Lambda}\over\lambda(\Lambda)$ so that
the coupling constant is now dimensionless. The regularized partition function
can now be formally written as,
\beq {\cal Z}[g, \Lambda, {\lambda(\Lambda)}]=\int {\cal D}_{\Lambda}[\phi]
{\cal D}_{\Lambda}[\sigma]
e^{-\int {d^3x
{\surd g}[{1\over 2}{\phi^i}(-\square_g+{\sigma}) {\phi_i}
-{{\sigma}\Lambda \over 2{\lambda(\Lambda)}}]}}\eeq
where ${\cal D}_{\Lambda}[\phi]=\prod\limits_{|k|<{\Lambda}} d\phi (k)$ and
similarly ${\cal D}_{\Lambda}[\sigma]$.
\vskip0.1in
\n {\bf The large $N$ limit}

Since it not possible to solve this theory exactly, we have to use some
approximation methods.
At this point there are a few different approximation methods
one could use in studying this
problem. These are the $2+\epsilon$ (or $4-\epsilon$)  expansion
methods, or the large $N$ expansion. We choose to study
the problem in the large $N$ limit.
In this paper we do all our analysis
to the leading order in the large $N$ expansion.

In the large $N$ limit, by which we mean $N\rightarrow \infty$ keeping
$N {\lambda(\Lambda)}$ fixed, the generating functional can be calculated
using the saddle point approximation. Also, we will assume that the number $n$
of components of the field $\phi_i$ that have non--zero  values in the ground
state
remains finite in this limit.
We will therefore redefine
$(N-n) \lambda(\Lambda)$ in the action as $\lambda(\Lambda)$ which will
remain fixed as $N\rightarrow \infty$. We also rescale the fields $\phi_i$ to
${\sqrt {N-n}}\phi_{i}$ for $i=1,2 \cdots n$.
We can now integrate out the first $N-n$ components of the $\phi$ field and
rewrite $\cal Z$ as,
\beq {\cal Z}[g,\Lambda,\lambda(\Lambda)]=\int{\cal D}_{\Lambda}[\phi]
{\cal D}_{\Lambda}[\sigma]
e^{{-{(N-n)\over 2}\bigl[{\rm TrLog}_{\Lambda} (-\square_g+ {\sigma})
+\int d^3x {\surd g}
[\sum\limits_{i=1}^n{\phi^i}(-\square_g+{\sigma}) {\phi_i}
-{\Lambda \over {\lambda(\Lambda)}}{\sigma(x)}]\bigr]}}
\eeq
We use the  $O(N)$ symmetry to rotate the unintegrated components to the
subspace indexed by $i=1\cdots n$ and denote these components by $b_i(x)$.

We first discuss the general case of the $O(N)$ sigma model on
a manifold with an arbitrary metric.
 This is given by the following set of equations
(``gap equations'') obtained by extremizing
the action with respect to
$b_{i}(x)$ keeping $\sigma(x)$ fixed and vice--versa.
\beq (-\square_g+\sigma) b_i=0, \eeq \label{qua}
\beq
\sum_{i=1}^{n} b_i^{2}= {\Lambda \over {\lambda(\Lambda)}}
- G_{\Lambda}(x, x; \sigma,g),\eeq \label{cin}
where $G_{\Lambda}(x, x; \sigma,g)
=\langle x|(-\square_g+ \sigma)^{-1}|x \rangle_
{\Lambda}$. $G(x^{\prime},x;\sigma,g)$ is the
two point correlation function of the $\phi$ fields.

Once we find the saddle point solutions of the action, we can compute the
free energy density, $W$, of the system at the saddle point,
to the leading order in the ${1 \over N}$ expansion.
\beq {W[g, \Lambda,\lambda(\Lambda)]}
={N \over 2}[{\rm TrLog}_\Lambda(-\square_g+\sigma)-\int d^3x {\surd g}
{\sigma} {\Lambda \over \lambda(\Lambda)}]. \eeq

The gap equations are the equations of motion of a classical  field theory: the
large $N$ limit of the $O(N)$ sigma model. The ground state will be given by
the solution with the least free energy. Other solutions representing solitons
or instantons of the ${1\over N}$ expansion are also of interest. If the
background  metric is homogenous, we expect the ground state solution of the
gap equations to be
constant. Then we can assume ( using $O(N)$ invariance) that just one component
of $b_i$ ( say $b_1=b$) is non--zero: this  has the meaning of spontaneous
magnetization.
 If $\sigma=m^2$ is positive, $m$ can be viewed as the mass of the scalar field
fluctuations around this background magnetization.

However, this interpretation
need not always make sense: stability of the ground state only demands that
$-\square_g+\sigma$ be positive, which does not always imply that $\sigma$
itself be positive.
The correlation function for fluctuations around the background $b_i$ is
the inverse of the operator $-\square +\sigma$. The condition of stability thus
coincides with the condition that these correlations not increase with
distance.
 The correlation length of the system should be
defined as $1\over \surd \lambda$, $\lambda$ being the smallest eigenvalue of
the operator $-\square_g+\sigma$.  In flat space this is the same as $m^{-1}$,
but in curved spaces, in general this is not the case.

\n {\bf \sect. Flat space $(R^3)$}

We will see in the next section that the short distance divergences of
the Green's function, $G_{\Lambda}(x, x; m^2,g)$,
are independent of the curvature of the space. It is
therefore natural to first study our theory on $R^3$.
Though this is a very well studied
case, \[polyabook], \[rosenstein],
let us briefly recapitulate the study of the fixed points in this theory.
On $R^3$, the Ricci scalar is zero.
Therefore the saddle point solutions are given by the
gap equations,
\beq m^2 b=0 \eeq \label{gapr31}
\beq
b^{2}= {\Lambda \over {\lambda(\Lambda)}}
- G_{\Lambda}(x, x; m^2,g)\eeq \label{gapr32}
\n In flat space, $-\square_{R^3}=-(\partial_x^2+ \partial_y^2+\partial_z^2)$.
The spectrum of the conformal laplacian is ${\rm Sp}(-\square_{R^3})=k^2$ where
$\vec k$ is a vector in $R^3$.
Therefore, $$G_{\Lambda}(x, x; m^2,g)={1 \over (2 \pi)^3}
{\int\limits^{\Lambda} {d^3k\over (k^2+m^2)}}$$
The gap equation \eqn{gapr31} admits 3 possible solutions with
$m^2=0, b\not=0$ or $m^2\not=0, b=0$ or both $m^2$ and $b$ vanish.

\n (1) When $m^2=0$ and $b\not=0$,
$$b^{2}={\Lambda \over {\lambda(\Lambda)}}
-{{1\over (2 \pi)^{3}}}
\int\limits^\Lambda {d^3k \over k^2}$$
At the critical coupling $\lambda_c(\Lambda)$ given by
$${\Lambda \over \lambda_c(\Lambda)}
={1\over (2 \pi)^3} \int\limits^\Lambda
{d^3k \over k^2}, $$
$b^2={\Lambda\over \lambda(\Lambda)}
-{\Lambda\over \lambda_c(\Lambda)}$ vanishes.

\n (2) When $m^2\not=0$ and $b=0$,
$$\eqalign
{{\Lambda \over \lambda_c(\Lambda)}-{\Lambda \over \lambda(\Lambda)}&=
{m^2 \over (2 \pi)^3} \int\limits^{\Lambda} {d^3k \over k^2 (k^2+m^2)}\cr
&={m \over 4 \pi}\cr}$$
from which it is seen that $m>0$
only when $\lambda > \lambda_c$. This corresponds to the unbroken phase of
the $O(N)$ symmetry. When
$\lambda < \lambda_c$, $m<0$ which is unphysical. If we do a more careful
analysis as in \[polyabook], we will be able to see that $m=0$ when
$\lambda < \lambda_c$. This is the broken phase where the $O(N)$ symmetry is
spontaneously broken and has given rise to $N-1$ Goldstone bosons with
$m=0$. At $\lambda=\lambda_c$, which separates the two phases,
$m$ goes to zero. This is the non--trivial fixed point of the theory, the
trivial fixed point being $\lambda_c=0$. We therefore see that at
$m=0$ and $b=0$, we have the critical theory at the non--trivial fixed point.
This fixed point is UV stable.

\n {\bf \sect. Short distance divergence structure of the Green's function}

We would next like to study our theory on curved 3--dimensional
manifolds $M$.
Before we
proceed, we would like to establish that the only divergences that arise in
the Green's function, in the gap equation in curved space, are those which
arise on $R^3$.
\beq {G_{\Lambda}(x, x; \sigma,g)
={\langle x|(-\square_g+ \sigma)^{-1}|x \rangle}_{\Lambda}
=\int\limits_{1\over \Lambda^2}^{\infty} dt
{\langle x|e^{-(-\square_g+\sigma)t}|x \rangle}} \eeq
Let $\lambda_n$ and $\psi_n(x)$ be the eigenvalues and eigenvectors
of $-\square_g+\sigma$. The Green's function can be written as follows:
$$G_{\Lambda}(x, x; \sigma,g)=\int\limits_{1\over \Lambda^2}^{\infty} dt
\quad h(t;x,x) $$
where $h(t;x,x)$ is the heat kernel of the operator
$-\square_g+\sigma$ and has the formal expansion
$$h(t;x,x)=\sum_n e^{- \lambda_n t} \psi_n^\ast(x) \psi_n(x).$$
The divergence in the Green's function comes from short distance and hence
from small $t$ region. Let us therefore isolate the divergent part,
from the finite part in the Green's function. This can be done by invoking the
 asymptotic expansion of the heat kernel, \[davies] which,
in $d$ dimensions, is given by
\beq h(t;x,y)\sim {e^{-{l^2(x,y) \over 4 t}} \over (4 \pi t)^{d \over 2}}
\sum_{n=0}^\infty a_n(x,y) t^n ,\eeq
 where $l(x,y)$ is the Riemannian distance
 between points $x$ and $y$ on the manifold $M$. The short distance behaviour
of the Green's function  is determined by the short time behaviour of the heat
kernel.  In particular, the divergence in $G_\Lambda(x,x,\sigma,g)$
is determined
by the leading terms in the short time expansion of the heat kernel.
If we formally substitute the asymptotic expansion of the heat kernel for
$d=3$, we find
that the leading term, with $a_{0}(x)=1$, is the only term that contributes to
the divergence in the Green's function:
$$G_{\Lambda}(x, x; \sigma,g)={\Lambda \over 4(\pi)^{3\over 2}} +
{\rm finite \quad part}.$$
We see that the divergent part of the Green's function is independent
of the metric $g_{\mu \nu}(x)$. Thus the critical value of the coupling
constant is independent of the metric. This argument is not quite
mathematically rigorous, since the expansion we are using is only asymptotic.
However,  this method has been used successfully in many renormalization
calculations and we believe it can be made rigorous using bounds on the short
distance behaviour of the heat kernel.

We will proceed to
study the $O(N)$ sigma model on some curved manifolds at
the critical point
$\lambda_c(\Lambda)$.  We find that at this critical point the free energy is
finite, confirming the above argument.

\n {\bf \sect. Zeta function regularization}

Although regularizations such as the Pauli-Villars regularization are easier
to understand physically, the zeta function
regularization is more
tractable when we are in curved space. We will thus be using this
regularization in our work. Since the critical value of the coupling constant
at which the theory becomes finite is independent of the background metric, we
compute this critical coupling on $R^3$.
The Green's function, $G(x,x;m^2,g)$, is regularized as,
$$G_s(x,x;m^2,g)= \langle x|(-\square_g+m^2)^{-s}|x \rangle
=\zeta_g(s,x).$$
$\zeta_g(s,x)$ is the `local zeta function',
$$\zeta_g(s,x)=\sum_{\lambda_n \not=0} |\lambda_n|^{-s} |\psi_n(x)|^2$$
where $\{\psi_n(x)\}$ is an orthonormal basis in the $\lambda_n$-eigenspace;
$\lambda_n$'s being
the eigenvalues of $(-\square_g+m^2)$ and the sum includes degeneracies.
If the eigenvalues are continuous the
 sum gets replaced by an integral.
Then, \beq G(x,x;m^2,g)=\lim_{s \rightarrow 1} \zeta_g(s,x).\eeq
(On homogeneous spaces such as the ones we will be considering in this paper,
$\zeta(s,x)$ turns out to be independent of $x$.)
The gap equation on $R^3$ in this regularization is
$$\lim_{s\rightarrow 1}[{1 \over {\lambda (s)}}
=b^2+ \int {d^3k \over (2 \pi)^3 (k^2+m^2)^s}
=b^2+{1\over 2 {\pi}^2}
\int\limits_0^\infty dt
{t^{s-1} \over \Gamma(s)} {\int k^2 dk e^{-(k^2+m^2)t}}], $$
where the regularized coupling $\Lambda \over \lambda(\Lambda)$ in the
Pauli-Villars regularization has been replaced by
$1 \over \lambda(s)$ in the zeta function regularization.
Here we have used the Mellin transform to analytically continue the
zeta function. Note that
 $$\zeta (s,x)=\int {d^3k \over (2 \pi)^3 (k^2+m^2)^s}$$
has no pole at $s=1$. (The local zeta function is seen to be independent
of $x$ due to the homogeneity of the space $R^3$).
It is now easy to verify that,
$$\lim_{s\rightarrow 1} [{1 \over {\lambda (s)}}
=b^2-{m^{3-2s} \over {(4 \pi)}^{ 3 \over 2}}
{\Gamma(s-{3 \over 2})\over {\Gamma (s)}}]. $$
Recall, by analytic continuation of the gamma function, we can see that
$\Gamma (-{1 \over 2})=-2 \surd \pi$ and thus,
$$\lim\limits_{s\rightarrow 1} {1 \over {\lambda (s)}}=b^2+{m \over 4 \pi}.$$
We have seen though from the analysis in the case of $M=R^3$ that at the
critical point, $m=0$ and $b=0$. $m$ and $b$ are physical quantities and
 are regularization independent.
Therefore in the zeta function regularization,
$\lim\limits_{s\rightarrow 1} {1 \over {\lambda_c(s)}}=0$.
We will be using this value of the critical coupling in all our future
calculations.

The critical coupling, while independent of the background metric, does depend
on the regularization scheme. In the
$\zeta$-function regularization, it is defined
through some analytic continuations and $1\over \lambda_c$ has the value zero.
This means that  one of the phases of the theory would correspond to negative
values of the coupling constant. There is no contradiction here, this is just a
peculiarity of this regularization method. Physical quantities such as free
energy still make sense.

In the zeta function regularization,
the free energy density of the $O(N)$ sigma model
in the large $N$ limit has a rather simple form:
\beq W(g)= {N \over 2} {\rm TrLog}(-\square_g+m^2).\eeq
Let us now consider the conformal invariance of the free energy density
in the case of
a general curved space. Temporarily we consider the situation in arbitrary
dimension $d$, which will make clear the special situation in an odd dimension
such as three ( i.e., there is no conformal anomaly).

If $g_{ij}\to e^{2f}g_{ij}$  and $\phi\to e^{(1-{d\over 2})f}\phi$, the action
$-\int\surd g d^dx \phi\square_g \phi$ is invariant. If we assign the
transformation $\sigma\to e^{-2f}\sigma$, the action $\int \phi[-\square_g
+\sigma]\phi \surd g d^dx$ is still conformally invariant. We can define a
determinant for this operator by
$${\rm det} [-\square_g+\sigma]=\cases {0, &if ${\rm
dim(ker}[-\square_g+\sigma])\not=0$\cr
(-1)^v e^{-\zeta^{\prime}(0)}, &if ${\rm dim(ker}[-\square_g+\sigma])=0$\cr}$$
where $v$ is the number of negative eigenvalues ( assumed to be finite in
number) of $-\square_g+\sigma$. Also, the zeta function,
$\zeta(s)=\sum_{\lambda_n\neq
0}|\lambda_n|^{-s}$, $\lambda_n$ being the eigenvalues of $-\square_g+\sigma$.

\n {\bf Theorem.} ~~ Given a $d$ dimensional compact manifold $M$,
${\rm det}(-\square_g+\sigma)$ is
conformally invariant under the transformation of the metric $g_{\mu
\nu}\rightarrow e^{2 f(x)} g_{\mu \nu}$ and $\sigma\to e^{-2f}\sigma$,
if $M$ is odd dimensional. If $M$
is even dimensional,
$$\delta_f {\rm det}[-\square_g+\sigma]=-(4 \pi)^{d \over 2} {\rm
det}[-\square_g+\sigma] \int_M 2 f(x) {\rm tr}a_{d \over 2}(x).$$
$a_{d \over 2}(x)$ is one of the coefficients in the asymptotic
expansion of the heat kernel $h(t; x,y)$
 which is a solution of the
diffusion equation,
$$(\partial_t-\square_g+\sigma)h(t; x,y)=0, \quad t>0;\quad
h(0;x,y)=\delta(x,y).$$
The asymptotic expansion of $h(t; x,y)$, as we have seen before, is,
$$h(t; x,y) \sim {e^{-{l^2(x,y) \over 4t}} \over (4 \pi t)^{d \over 2}}
\sum_{k=0}^\infty a_k(x,y) t^k$$ where $l(x,y)$ is the distance between
the points $x$ and $y$ on M.

\n A proof of this theorem in the case $\sigma=0$, in the zeta function
formulation is given in
\[parker]. This proof can be extended without much difficulty to
our case where the operator is $-\square_g+\sigma(x)$.  The explicit expression
 for $a_{k}(x)$ will change, but this is not of interest in the odd dimensional
case.

Rather than grind through the proof with the minimal changes required to Ref.
\[parker] we will give a simple ( but not rigorous) argument which we believe
captures the essence of the conformal anomaly. Under an infinitesimal conformal
transformation, the operator $-\square_g+\sigma$ transforms like $\sigma$
itself: a density of weight $-2$.  The variation of the $\zeta$-function is,
\beq
\delta\Tr[-\square_g+\sigma]^{-s}=
-s\Tr[-\square_g+\sigma]^{-s-1}\delta[-\square_g+\sigma]=
  (2s)\Tr[-\square_g+\sigma]^{-s}\delta f.
\eeq
Differentiating the two sides with respect to $s$ and setting $s=0$, we get
\beq
      \delta\log\det[-\square_g+\sigma]=-\delta \zeta'(0)=
	-2\lim_{s\to 0}\Tr[-\square_g+\sigma]^{-s}\delta f.
\eeq
Although this statement is in essence true, to  justify  the limits,
some more  work is necessary.

  Formally the r.h.s. is the
trace of the identity operator weighted by $\delta f$. This can be evaluated
by putting in the asymptotic form of the heat kernel and performing a Mellin
transform.   The only term which will contribute ( in the limit) is $a_{d\over
2}$, as can be verified easily. These steps are common in the computation of
anomalies in the physics literature. The proofs in \[parker] are somewhat more
precise.

To summarize, the large $N$ limit of the $O(N)$ sigma model at the fixed
point
can be viewed as a classical field theory with action
\beq
	\Gamma[b_i,\sigma]=\half \int b_i[-\square_g +\sigma] b_i\surd g d^3 x
+\half\log\det[-\square_g+\sigma].
\eeq
The gap equations are the Euler--Lagrange equations of this non--local action.
This action is conformally invariant under the above transformations. In this
sense the large $N$ limit of the $O(N)$ sigma model is a conformal field
theory.  The field  $b_i$  is a primary field in the sense that its correlation
functions are conformally covariant.
\vskip0.1in
We are now ready to study specific examples.
The next two special cases we investigate are the $O(N)$ sigma model on
the manifolds, $R^2 \times S^1$ and $S^1 \times S^1 \times R$, both
of zero Riemann curvature. They are however not conformally equivalent
to $R^3$. We will see that in these cases $m$ is non-zero at the critical
point. In these cases, $m$ is the inverse correlation length and therefore
the correlation length is
finite at criticality. This is due to the finite size of the
manifold in some directions. More generally, whenever the manifold is
not conformally equivalent to $R^3$, we should expect the value of $m$
at criticality to be non-zero. Our computations confirm this conjecture.

\n {\bf \sect. Finite size effects on $R^2\times S_{\beta}^1$}

This case has been well studied in the papers \[sachdev], \[rosenstein]
in the
Pauli-Villars regularization. We will study it in the zeta function
regularization \[fradkin]
to make later comparisons easier. Also, the calculation is
technically simpler in the zeta function regularization.

On $R^2 \times S^1$ we choose the metric tensor to be
$g_{\mu \nu}=(1,1,{\beta}^2)$.
The radius of the circle, $\beta$, can be thought of as
inverse temperature for a system on the 2--dimensional flat space $R^2$.
The conformal laplacian on this space is
$$-\square_g=-(\partial_x^2+\partial_y^2+{1\over {\beta}^2}
\partial_\theta^2),$$
where $(x,y)$ are co-ordinates in $R^2$ and $\theta$ is the
local co-ordinate on $S^1$. The spectrum of $-\square_g$ is,
$$Sp(-\square_g)=k^2+ {4 \pi^2 n^2\over{\beta}^2}, $$
$n=0,\pm 1, \pm 2, \cdots$.
The gap equations are
\beq m^2 b=0 \eeq \label{gapr2s11}
\beq\lim_{s\rightarrow 1} [ b^2={1 \over {\lambda (s)}}-{1 \over \beta}
\sum\limits_n
\int {d^2k \over (2 \pi)^2 (k^2+{4 \pi^2 n^2\over{\beta}^2}+m^2)^s}].
\eeq \label{gapr2s12}

\noindent At the critical point $\lim\limits_{s \rightarrow 1}
{1 \over \lambda(s)}=0$.
Taking the Mellin transform of the above equation, we get,
$$-\lim_{s\rightarrow 1}
{1 \over \beta} \int\limits_0^\infty dt {t^{s-1}\over \Gamma(s)}\sum\limits_n
\int\limits_0^\infty {dk\over 2 \pi} k
{e^{-(k^2+{4 \pi^2 n^2\over{\beta}^2}+m^2)t}}=b^2.$$
We integrate over $k$ and use the Poisson sum formula to separate out the
divergent part in the small $t$ region of the sum so that we are now able
to interchange
the sum and the integral. The Poisson sum formula in this case is,
\beq\sum\limits_n e^{-({4 \pi^2 n^2\over{\beta}^2})t}
= {{\beta}\over {(4 \pi t)^{1 \over 2}}}\sum_n e^{-n^2 {\beta}^2\over {4 t}}
.\eeq \label{poisson1}
\n On using this formula, the gap equation simplifies to,
$$-\lim_{s\rightarrow 1}\Bigl [
\sum\limits_{n=1}^\infty {2 \over (4 \pi)^{3 \over 2}} \int\limits_0^\infty dt
{t^{s-{5 \over 2}}\over \Gamma(s)} e^{-(m^2 t+{{n^2 \beta^2}\over {4 t}})}
+{1 \over (4 \pi)^{3 \over 2}} \int\limits_0^\infty dt
{t^{s-{5 \over 2}}\over \Gamma(s)} e^{-m^2 t}\Bigr ]=b^2.$$
We use the standard integral,
\beq\int\limits_0^\infty dt\quad t^{\nu -1} e^{-({\sigma \over t}+{\gamma t})}
= 2 {\sigma \overwithdelims () \gamma}^{\nu \over 2}
K_{\nu} (2 {\sqrt {\sigma\gamma}}) ;\quad {\rm Re}{\sigma} > 0,
{\rm Re}{\gamma} >0,\eeq \label{besselint}
\noindent where $K_{\nu}$ is
the MacDonald's function.
The gap equation is then:
$$-\lim_{s\rightarrow 1}
\Bigl [\sum\limits_{n=1}^\infty{4 \over (4 \pi)^{3 \over 2}}
{n^2 {\beta}^2 \overwithdelims () 4 m^2}^{{s \over 2}-{3 \over 4}}
K_{s-{3\over 2}}(m \beta n)+ {m^{3-2s} \over (4 \pi)^{3 \over 2}}
{\Gamma(s-{3 \over 2})} \Bigr ]=b^2$$
Noting that, $K_{-{1 \over 2}} (m \beta n)=K_{1 \over 2} (m \beta n)=
{{\pi \overwithdelims () 2 m \beta n}^{1\over 2}} e^{-m \beta n}$,
in the limit $s \rightarrow 1$
the gap equation reduces to a very simple equation:
\beq{1 \over 2 \pi \beta}{\rm Log}2 \sinh{m \beta \overwithdelims ()
2}=b^2.\eeq
We know from equation \eqn{gapr2s11} that either $m=0$ or $b=0$ or both.
The equation above is not satisfied for $m=0$ since $b^2$ is always
positive. Hence we require $b=0$ to satisfy equation \eqn{gapr2s11}.
This gives us $m={2 T {\rm Log}\tau}$, as in \[sachdev],
where $\tau$ is the Golden mean
$(1+ \surd 5)\over 2$ and $T={1 \over \beta}$ and can be interpreted as the
temperature of the system.
\vfill
\eject
\n {\bf Free energy density on $R^2 \times S_{\beta}^1$}

The
regularized free energy density, at the critical point,
on $R^2 \times S^1$ is,
$${W_c(\beta)\over N}={1\over 2}{\rm TrLog}(-\square_g+m^2)=-{1\over 2}
\zeta_{R^2 \times S^1}^{\prime}(0).$$
More explicitly,
\beq {W(\beta)\over N}=-\lim_{s\rightarrow 0} {d\over {ds}} {1\over 2 \beta}
\sum\limits_n \int {d^2k \over (2 \pi)^2
(k^2+{4 \pi^2 n^2\over{\beta}^2}+m^2)^s}.\eeq
We Mellin transform the r.h.s and simplify the above expression as before to,
$${W(\beta)\over N}={2 \over (4 \pi)^{3 \over 2}}[-{\surd \pi \over 3} m^3
- {2 m \overwithdelims () \beta}^{3 \over 2}
\sum\limits_{n=1}^\infty  {1 \over n^{3 \over 2}}
K_{3\over 2}(m \beta n)].$$
Recalling, $K_{3\over2}(x)
=-2 {d \over dx} K_{1 \over 2} (x)-K_{1\over 2}(x)$, we obtain,
\beq {W(\beta)\over N}=-{1 \over (4 \pi)} [{m^3 \over 3}+
{2 m \overwithdelims () \beta^2}
\sum\limits_{n=1}^\infty {e^{-(m \beta n)} \over n^2}
+{2 \overwithdelims () \beta^3}
\sum\limits_{n=1}^\infty {e^{-(m \beta n)} \over n^3}].\eeq
We know though that at the critical point, $m \beta = \log{\tau}^2
=-\log(2-\tau)$. We also recall the power series representation of the
polylogarithm, $\sum\limits_{n=1}^\infty {x^n \over n^p}={\rm Li}_p(x)$.
Putting all these together, the free energy density, in agreement with
\[sachdev], is
\beq {W(\beta)\over N}={1 \over (2 \pi {\beta}^3)} [{1 \over 6} \log^3(2-\tau)
+\log(2-\tau){\rm Li}_2(2-\tau)-{\rm Li}_3(2-\tau)]
=-{2 \over {5 \pi {\beta}^3}} {\rm Li}_3(1)\eeq
The last equality can be arrived at by using polylogarithm identities and is
derived in Ref.\[sachdev].
Using the expression for the regularized
free energy density obtained from hyperscaling for
a d--dimensional slab geometry\[cardy],
$${W(\beta) \over N}=
-{\Gamma({d \over 2}) \zeta(d)\over {\pi}^{d \over 2} {\beta}^3} \tilde c, $$
it is seen that, $\tilde c={4 \over 5}$ (a rational number).
(Note, here $\zeta(d)$ is
 the Riemann zeta function.)
It is therefore
of some mathematical interest to study the free energy of the large
$N$--limit of the $O(N)\;\sigma$ model. There is a possibility that
this is a
three dimensional analogue of rational conformal field theory. We hope that the
explicit calculations described here will be useful to test this conjecture.

\n {\bf \sect. Study of the $O(N)$ sigma model on manifolds of the type
$\Sigma \times R$}

In the next three subsections we will be studying the $O(N)$ sigma model on
a manifold $M=\Sigma \times R$. We therefore discuss the general form of
the gap equations on such manifolds before we consider specific cases of
$\Sigma$.

The Green's function we need can be written in terms of the geometry of the
2-manifold $\Sigma$:
\beq \eqalign{G_{s}(x, x; m^2,g)&=\langle x|(-\square_g+ m^2)^{-s}|x \rangle
\cr
&=\int\limits_0^\infty {dt \over \Gamma(s)}
t^{s-1} \langle x| e^{-[-\nabla_{\Sigma}^2+{\xi R}-\partial_u^2+m^2]t}|x
\rangle\cr
&={1 \over {\sqrt{4 \pi}}}
\int\limits_0^\infty {dt \over \Gamma(s)}
t^{s-{3 \over 2}} \langle x|
e^{-[-\nabla_{\Sigma}^2+{\xi R}+m^2]t}|x \rangle \cr
&={1 \over {\sqrt {4 \pi}}} {\Gamma(s-{1 \over 2}) \over \Gamma(s)}
\zeta_{\Sigma}(s-{1 \over 2},x,m^2)\cr}\eeq
Here,
$$
\zeta_{\Sigma}(s,x,m^2)=\langle x|[-\nabla_{\Sigma}^2+{\xi R}+m^2]^{-s}|x
\rangle.$$
At the critical point, the gap equations will then be of the form,
\beq (-\square_g+m^2(x)) b_i(x)=0\eeq
\beq
\sum_i b_i(x)^2=-\half\zeta_\Sigma(\half,x,m^2).
\eeq
The zeta function on the r.h.s. is analytic at $s=\half$ so that this
equation is finite (the singularities of the zeta function of an even
dimensional  manifold occur at negative integer values of $s$).

To evaluate
$\zeta_{\Sigma}(s-{1 \over 2},x,m^2)$, we have to find the spectrum of the
conformal laplacian, $-\nabla_{\Sigma}^2+\xi R+m^2$, on the space $\Sigma$.
The specific examples we are going to consider are the ones with
$\Sigma$ having zero, constant
positive and constant negative curvature.
In this case we can look for solutions of the gap equations with
$m^2$ and $b_i$ constant;
then $b_i$ can be chosen to have only one non--zero component.
\vfill
\eject
\n {\bf \subsect. $S^1 \times S^1 \times R$}

For simplicity let us study the case where the radii of the two
circles are the same and are denoted by $\rho$.
This is a space of zero curvature. The Ricci scalar $R=0$ and
the conformal laplacian is just the ordinary laplacian and is given by
$$-\square_g=-{1\over \rho^2}
(\partial_\theta^2+\partial_\phi^2)+\partial_z^2,$$ where
$\theta$ and $\phi$ are local co--ordinates on the two circles and $z$ the
co--ordinate on $R$. The spectrum of the conformal laplacian is,
$${\rm Sp}(-\square_g)={4 \pi^2 \over \rho^2}(p^2+q^2)+k^2$$ where,
$p,q=0, \pm 1, \pm2, \cdots$ and $k$ takes values on the real line.
The gap equations on $S^1 \times S^1 \times R$ are,
\beq m^2 b=0 \eeq \label{gaps1s1r1}
\beq\lim_{s \rightarrow 1}\Bigl[{1 \over \lambda(s)}
-{1 \over 2}
\zeta_{S_\rho^1 \times S_\rho^1}(s-{1 \over 2},x)=b^2 \Bigr]
\eeq \label{gaps1s1r2}
At the critical point,
$\lim\limits_{s \rightarrow 1} {1 \over \lambda_c(s)}=0$.
We use the integral representation of the zeta function,
$$\zeta_{S_\rho^1 \times S_\rho^1}(s-{1 \over 2},x)
={1 \over \rho^2} \int\limits_0^\infty dt \quad
{t^{s-{3 \over 2}} \over \Gamma(s-{1 \over 2})} \sum_{p,q}
e^{-{4 \pi^2 \over \rho^2}(p^2+q^2)t-m^2 t}$$
and after performing calculations similar to that in the example of
$R^2 \times S^1$,
at the critical point, \eqn{gaps1s1r2} simplifies to:
\beq -{1 \over 4}-{1 \over m \rho}{\rm Log}(1-e^{-m \rho})
+ \sum_{p,q=1}^\infty {{e^{-{\sqrt (p^2+q^2)} m \rho}} \over m \rho
{\sqrt {p^2+q^2}}}=-{ \pi b^2 \over m} \eeq \label{doublesum}
\n This equation is difficult to solve as the double sum we are left with
is not an obvious one. We can see that $m\not=0$ without actually solving
the equation. In essence, we
put in an ansatz $m \rightarrow 0$ and we will show that this is
inconsistent with the gap equation and hence $m\not=0$ in this case
at the critical point. If $m$ were small, we could approximate the double
sum by a double integral and the equation \eqn{doublesum} can be written as,
$$-{1 \over 4} -{1 \over m \rho} {\rm Log}(1-e^{-m \rho})
+\int\limits_0^\infty dp \int\limits_0^\infty dq \quad
{{e^{-{\sqrt (p^2+q^2)} m \rho}} \over m \rho
{{\sqrt (p^2+q^2)}}}=-{ \pi b^2 \over m}$$
The integral in the above expression can be easily performed using polar
co-ordinates and is evaluated to be ${ \pi \over {2 (m \rho)^2}}.$
Also, when $m \rightarrow 0$, we can approximate ${\rm Log}(1-e^{-m \rho})$
to ${\rm Log}{m \rho}$ and the above equation reduces to a transcendental
equation for $m \rho$,
\beq {(m \rho)^2 \over 4} +(m \rho) {\rm Log}{m \rho}
-{\pi \over 2}={ \pi b^2 m \rho^2}.\eeq

\n It is immediately apparent from the above equation that $m=0$ cannot be one
of its solutions. This implies that $b$ has to be zero at the
critical point in order to satisfy the gap equation \eqn{gaps1s1r1}.
Hence the solutions to the gap equations in the
case of $M=S^1 \times S^1\times R$ are $m\not=0$ and $b=0$ as we expected
and $m$ is the solution to \eqn{doublesum}. Since $m\not=0$ here, the smallest
 eigenvalue of $-\square_g+m^2$ is non-zero and therefore the
correlation function
will decay exponentially.


\n Once we solve for $m$, the free energy density can be computed at the
 critical point with this value of $m$.

\n {\bf \subsect. $S_\rho^2 \times R$ - example of
a space of constant positive curvature}

Since the space
$S^2$ with radius $\rho$ has finite volume, we might tend to expect $m$
to be non-zero at criticality. But we will see that
this is not true; the manifold $S^2 \times R$ is conformal to $R^3-\{0\}$
and it turns out that in fact $m=0$. Thus we see that what matters,
for $m$ to be zero or otherwise, is the
conformal class of the metric and not the `size' of the system.
$m=0$ does not however mean an infinite correlation length on $S^2 \times R$;
the correlation length at criticality is in fact finite.

In order to find the conformal laplacian on $S^2 \times R$, we
need to calculate the Ricci scalar on $S^2$. This is a standard calculation
and will in the end yield
$R={2 \over \rho^2}$ and therefore $\xi R={1 \over 4 \rho^2}$.
Since conformal
curvature plays an important role in our study of the critical theory,
it is worthwhile to demonstrate this equivalence of $S^2 \times R$ to
$R^3-\{0\}$.

\n Let the metric on
$S_\rho^2 \times R$ be denoted by $g_1$ and that on $R^3-\{0\}$ by $g$.
The line element on $S_\rho^2 \times R$ is,
$$ds_{S^2 \times R}^2=\rho^2 (du^2+d\Omega^2)$$ where, $u$ is the co-ordinate
 on $R$ and
$d\Omega$ is the solid angle.
The line element on $R^3-\{0\}$ in spherical polar co-ordinates is,
$$ds_{R^3-\{0\}}^2=dr^2+r^2 d\Omega^2.$$
On writing $r$ as $r=\rho e^u$, the line element on $R^3-\{0\}$ becomes,
$$ds_{R^3-\{0\}}^2=\rho^2 e^{2u}(du^2+ d\Omega^2).$$
We immediately see that the metrics $g$ and $g_1$ are related by a
conformal transformation,
$g_1=e^{2 f} g$ with $f(x)=-u$.
Thus the manifolds $S^2 \times R$ and $R^3-\{0\}$
are conformally equivalent. We can now use this fact to fix $\xi R$ on
$S^2 \times R$.
For a conformal transformation, $g \rightarrow g_1=e^{2f} g$, the
scalar curvature term $\xi R$ transforms as follows \[parker]:
$$\xi R_1=e^{-(d+2){f \over 2}} \square_g e^{(d-2) {f \over 2}}.$$
Note, $\square_g=\nabla_g^2$ since the Ricci scalar is zero for $R^3-\{0\}$.
Also, in our case, $d=3$ and $f(x)$ is equal to $-u$.
Hence,
$$\xi R_1=e^{{5u \over 2}} \nabla_g^2 e^{-{u \over 2}}$$
where,
$$\nabla_g^2
= \rho^{-2} e^{-2u} (\partial_u^2+ \partial_\theta^2
+{1 \over \sin^2\theta} \partial_\phi^2).$$
This gives $\xi R_1={1 \over 4 \rho^2}$.
The conformal laplacian on $S^2 \times R$ is therefore given by
$$-\square_{S^2 \times R}=-\nabla_{S^2 \times R}^2+{1 \over 4 \rho^2}$$
with the spectrum as
\beq {\rm Sp}(-\square_{S^2 \times R})=[{(l+{1 \over 2})^2 \over \rho^2}+k^2],
\eeq
where $l=0,1,2,\cdots$
and $k \in R$ with degeneracy $(2l+1)$.
Notice that the conformal laplacian on $S^2 \times R$ has no zero modes.

\n The gap equations on $S^2 \times R$ are
\beq({1 \over 4 \rho^2}+m^2) b =0,\eeq \label{gaps2r1}
\beq{\lim_{s\rightarrow 1}}
[{1\over {\lambda(s)}}-{1 \over 2}
\zeta_{S_\rho^2}(s-{1 \over 2},x)= b^2].\eeq \label{gaps2r2}
\n From equation \eqn{gaps2r1} we see that $b=0$ since
$m^2+ {1 \over 4 \rho^2}$
cannot be equal to zero because both $m$ and $\rho$ are positive.
\beq \zeta_{S_\rho^2}(s-{1 \over 2},x)=
{1\over {\rho^2}}\sum_{l={1\over 2}}^\infty
{{2 l}\over ({l^2 \over \rho^2}+m^2)^{s-{1 \over 2}}}.\eeq
On using the Mellin transform of the above zeta function, at the critical
point, \eqn{gaps2r2} gives,
\beq {\lim_{s\rightarrow 1}}
\int\limits_0^\infty dt {t^{s-{3 \over 2}}\over \Gamma(s)} e^{- m^2 t}
\sum_{l={1\over 2}}^\infty 2 l e^{-{l^2 \over \rho^2}t}=0.\eeq
As before we see that the sum is divergent in the small $t$ region and we
therefore have to separate out the divergent piece in the sum before we can
interchange the sum and the integral over $t$.
To do this we need an analog of the Poisson sum formula for the
case of the sum over half-integers $l$ which turns out to be the following:
\beq {1 \over 2 \pi} \sum_{l={1\over 2}}^\infty 2 l e^{-{l^2 \over \rho^2}t}
={\rho^2 \over 4 \pi t}+{\rho^2\over (4 \pi t)^{3 \over 2}}
{\rm P} \int\limits_{-\infty}^\infty dx
({x \over 2 \rho} {\rm cosec}{x \over 2 \rho}-1) e^{-{x^2 \over 4 t}}
\eeq\label{poisson2}
where, by ${\rm P}\int\limits_{-\infty}^\infty dx
({x\over 2 \rho}{\rm cosec}{x \over 2 \rho}-1) e^{-{x^2 \over 4 t}}$
we mean the principal value of the integral;
$({x\over 2 \rho} {\rm cosec}{x \over 2 \rho}-1)$
has simple poles at all non-zero
integral multiples of $2 \pi \rho$.
We give a brief derivation of this formula in appendix B.
On using the Poisson sum formula \eqn{poisson2},
in the limit $s \rightarrow 1$, the gap equation reduces to,
\beq{\rho \over  {\sqrt \pi}}
{\rm P}\int\limits_{-\infty}^\infty dx \quad
({x} {\rm cosec}{x}-1) \int\limits_0^\infty dt \quad
t^{-2} e^{-{x^2 \rho^2 \over t}-m^2 t}
+m \Gamma(-{1 \over 2})=0.\eeq \label{solvem}
\n The integral over $t$ can be easily performed using \eqn{besselint} and will
give a Macdonald's function. But this way we find it hard to extract the
solution for $m$ from the expression we get. We expect $m=0$
to be the solution, though, since we showed that $S^2 \times R$ is conformally
equivalent to $R^3-\{0\}$. Let us therefore try putting the ansatz
$m=0$ as the solution in the l.h.s of the above gap equation and
see if this is a consistent solution.
On putting $m=0$ in the l.h.s of \eqn{solvem} and performing the integral
over $t$, we get,
$${\rm L.H.S}=
{1 \over  {\sqrt \pi} \rho} {\rm P}\int\limits_{-\infty}^\infty dx
{1\over x} ({\rm cosec}{x}-{1 \over x})$$
We can easily check that the above integral is zero,
thus giving us the l.h.s of the gap equation to be zero, consistent with
the r.h.s. Hence $m=0$ is indeed the correct solution to the gap equation
in this case. We therefore find that at the critical point, $m=0$ and
$b=0$ on $S^2 \times R$.

Although $m=0$ at criticality, the correlation length of
this system is not infinite. If we consider the correlation
function $<\phi_i(x,u)\phi_j(y,0)>$ as a function of
$x,y\in S^2$ and $u\in R$, it will decay like an
exponential in  $u$ as $u\to\infty$.
(In the $x,y$ directions the space has finite radius.)
This is because the operator $-\nabla_{\Sigma}^2+\xi R+m^2$ has no zero modes.
\vfill
\eject
\n {\bf Free energy density on $S^2 \times R$}

\n The regularized free energy density on $S^2 \times R$
is given by,
$${W(\rho)\over N}
=-\lim_{s \rightarrow 0} {d \over ds}{1 \over 4 \pi \rho^2}
\int\limits_{-\infty}^\infty dk \quad
\sum_{l={1 \over 2}}^\infty  {2l \over (k^2+{l^2 \over \rho^2}+m^2)^s}
-\lim_{s \rightarrow 1} \int d^3x {\surd g} {m^2 \over \lambda(s)}.$$
Using the analytic continuation of the zeta function, the Poisson sum formula
 \eqn{poisson2} and performing the
calculations which are fairly straightforward,
we get, at the critical point,
$${W_c(\rho)\over N}
={1 \over 16 \pi}
\bigl[{\rm P} \int\limits_{-\infty}^\infty dx
({x \over 2 \rho} {\rm cosec}{x \over 2 \rho}-1)
\int\limits_0^\infty dt {t {\log t} \over \Gamma(s) }
e^{-{x^2 \over 4 }t} -{\rm P} \int\limits_{-\infty}^\infty dx
{4 \overwithdelims () x^2}^{2}
({x \over 2 \rho} {\rm cosec}{x \over 2 \rho}-1) \bigr]$$
The first term vanishes since in the limit ${s \rightarrow 0}$,
$ \Gamma(s) \rightarrow {1 \over s} $. We can again perform the integral over
$x$ in the second term and it can be verified
that it is zero. We see therefore that the
regularized free energy density,
 \beq {W_c(\rho) \over N}=0\eeq on $S^2 \times R$.
This just means that the free energy
density on $S^2 \times R$ is the same as that on $R^3$ at the critical point
which is what we should expect from general considerations of conformal
equivalence of the spaces $S^2 \times R$ and $R^3-\{0\}$.

\n {\bf \subsect. $H^2 \times R$ - example of a space of constant negative
curvature}

Let us consider as an example of constant negative curvature $M=H^2 \times
R$, where $H^2$ is a two--dimensional hyperboloid.
Let the co-ordinate on $R$ be denoted by $u$ and the hyperboloid be
parameterized as
$$H^2=\{z=(x,y):x \in R, 0<y<\infty\}$$
with line element and laplacian given by
$$ds^2={\rho^2 \over y^2} (dx^2+dy^2) \quad {\rm and} \quad
\nabla_{H^2}^2={y^2 \over \rho^2} (\partial_x^2+\partial_y^2)$$
respectively.
The Ricci scalar on this space is $R=-{2 \over \rho^2}$. Therefore
$\xi R=-{1 \over 4 \rho^2}$. The gap equations are,
\beq (-{1 \over 4 \rho^2}+m^2) b=0 \eeq \label{gaph2r1}
\beq {\lim_{s \rightarrow 1}} [{1 \over \lambda(s)}
-{1 \over 2}
\zeta_{H^2}(s-{1 \over 2},x)]=b^2. \eeq \label{gaph2r2}
\n We have to now calculate $\zeta_{H^2}(s-{1 \over 2},x)$. At the expense of
wearying the reader, we give more calculational details in this example as
 it is qualitatively different from those we considered earlier.

\n To determine $\zeta_{H^2}(s-{1 \over 2},x)$, we need to find the spectrum of
the conformal laplacian, $-\nabla_{H^2}^2+\xi R$ on the hyperbolic space. This
spectrum is continuous and therefore we have to find the density of states
$\mu(\lambda)$. The object that is of interest to us is,
$$\zeta(s-{1 \over 2},x)=\langle x| [-\nabla_{H^2}^2-{1 \over 4 \rho^2}+m^2]
^{-(s-{1 \over 2})}|x \rangle =\int d\lambda \quad \mu(\lambda)
[\lambda-{1 \over 4 \rho^2}+m^2]^{-(s-{1 \over 2})}$$
(for simplicity, we assume that $\rho=1$). If we define $\nu$
by $\lambda=\nu (1-\nu)$,
the resolvent,
\beq R(\nu)=[-\nabla_{H^2}^2-\nu (1-\nu)]^{-1}\eeq is a single valued function
of
$\nu$.
This means that the resolvent has a
branch cut on the complex $\lambda$ plane with $\lambda={1 \over 4}$
as the branch point. The explicit form of the resolvent is given in \[lang]:
$$R(\nu) f(z)=\int\limits_{y>0} \phi_{\nu}(u(z,z^{\prime})) f(z^{\prime})
{dx^{\prime}dy^{\prime} \over {y^{\prime}}^2}$$
where, $z=(x,y)$, $u(z,z^{\prime})={|z-z^{\prime}|^2 \over 4 y y^{\prime}}$
and,
\beq\phi_{\nu}(u)={1 \over 4 \pi} \int\limits_0^1 dt
\quad [t(1-t)]^{\nu-1} (t+u)^{- \nu}.\eeq
If we consider the matrix elements,
$$\langle z|R(\nu)|z^{\prime} \rangle=\phi_{\nu}(u(z,z^{\prime}))$$
we can get the density of states
$\mu(\lambda)$ which is given by the discontinuity
of the resolvent of the Laplace operator across the branch cut,
\beq \mu(\lambda)={1 \over 2 \pi i}
[\langle z|R(\lambda+i\epsilon)|z^{\prime}\rangle
-\langle z|R(\lambda-i\epsilon)|z^{\prime} \rangle]_{z=z^{\prime}}\eeq
to be,
\beq\mu(\lambda)={1 \over 8 \pi}\Theta(-{1 \over 4}+\lambda)
\tanh{\pi {\sqrt{-{1 \over 4}+\lambda}}}.\eeq
We can now compute $\zeta_{H^2}(s-{1 \over 2},x)$.
$$\zeta_{H^2}(s-{1 \over 2},x)=\int\limits_0^\infty dt \quad {
t^{s-{3 \over 2}} \over \Gamma(s-{1 \over 2})}
\int\limits_{1 \over 4}^\infty d\lambda\quad
\tanh{\pi{\sqrt{-{1 \over 4}+\lambda}}} e^{-[{\lambda}-{1 \over 4}+m^2]t}$$
We change variables, $k^2={-{1 \over 4}+\lambda}$ and perform the integral
over $t$. On substituting the resultant expression for
$\zeta_{H^2} (s-{1\over 2},x)$ in the
 gap equation \eqn{gaph2r2}, we obtain, at the critical point,
\beq
[m +\int\limits_0^\infty dk \quad
{k \over {\sqrt{k^2+m^2}}} (1-\tanh{\pi k})]=b^2.\eeq \label{solveb}
\n We see that each term in
the l.h.s is manifestly positive . Therefore, $b=0$ cannot be
a solution when $m\not=0$. If $m=0$, again it is easy to check that the
l.h.s is non-zero (it is equal to $\ln2 \over \pi$)
and hence $b=0$ cannot satisfy the above equation.
Therefore, in order to satisfy the gap equation \eqn{gaph2r1}, we
require,
$$m={1 \over 2 \rho}$$
 (since we set $\rho=1$, $m={1 \over 2}$)
at the critical point.
We see that negative curvature has induced the symmetry to be
spontaneously broken at
the critical point and there is a non-zero spontaneous
magnetization given by the order parameter $b$.

The value $m={1\over 2\rho}$ is special in
that $[-\nabla_{\Sigma}^2+\xi R+m^2]b=0$ has  a
constant solution  for that value. The correlation functions decay like
a power law in the variable $u(z,z^{\prime})$,
but an exponential in the variable $d=\rho~{\rm
 arcosh}\;(1+2u)$ (which has the meaning of geodesic distance).

We can now evaluate the order parameter $b$ at criticality from
\eqn{solveb}.
We expand ${1 \over \sqrt {k^2+m^2}}$ binomially and simplify \eqn{solveb}
to the following:
$$ b^2=[m + 2 \int\limits_0^\infty dk \quad
\sum_{r=0}^\infty {{1 \over 2} \choose r}
{{k \overwithdelims () m}^{2r+1} \over e^{2 \pi k}+1}].$$
The integral over $k$ can be performed and and after restoring the appropriate
factors of $\rho$, we can express
$b$ at criticality as,
\beq b^2\rho ={{\sqrt \pi}\over 2} [1 +
\sum_{r=0}^\infty {{1 \over 2} \choose r}
{2\over \pi^{2r+2}} (2r+1)! (1-2^{-2r-1}) \zeta(2r+2)] \eeq \label{bseries}
(this sum can also be written as a similar sum over Bernoulli numbers).

\n {\bf Free energy density on $H^2 \times R$}

The free energy density on $H^2 \times R$ is given by,
\beq W(\rho)= {N \over 2} [{1 \over 2 \pi} \int dp \int d\lambda
\quad \mu(\lambda) \quad
{\rm Log}(p^2+{\lambda \over \rho^2}+{1 \over 4 \rho^2}+m^2)
-\lim_{s \rightarrow 1} \int d^3x {\surd g} {m^2 \over \lambda(s)}]\eeq
At the critical point, the regularized free energy density
is given by (we again set $\rho=1$),
$${W_c(\rho) \over N}= -\lim_{s \rightarrow 0} {d \over ds}\bigl[
{1 \over 4 \pi} \int dp
\int\limits_0^\infty dt \quad {t^{s-1} \over \Gamma(s)}
\int\limits_{1 \over 4}^\infty d\lambda
\quad {\tanh\pi \sqrt{{-1 \over 4}+ \lambda}}
e^{-{(p^2+\lambda-{1 \over 4}+m^2)}t}\bigr]$$
where $m={1 \over 2 \rho}$.

We first perform the integral over $p$ and change variables,
$k^2= -{1 \over 4}+\lambda$ and then
integrating over $t$ gives us the following expression for the free energy
density:
\beq {W_c(\rho) \over N}= -\lim_{s \rightarrow 0} {d \over ds}
{1 \over  4{\sqrt \pi}} \bigl[m^{3-2s} {\Gamma(s-{3 \over 2}) \over \Gamma(s)}
+2 {\Gamma(s-{1 \over 2}) \over \Gamma(s)}
\int\limits_0^\infty dk \quad k (m^2+k^2)^{{1 \over 2}-s}({\tanh\pi k}-1)
\bigr]
.\eeq
We once again use the binomial expansion of $(m^2+k^2)^{{1 \over 2}-s}$ and
 perform the integral over $k$ and obtain,
$$\eqalign{{W_c(\rho) \over N}= &-\lim_{s \rightarrow 0} {d \over ds}
{1 \over  4{\sqrt \pi}}\bigl[m^{3-2s} {\Gamma(s-{3 \over 2}) \over
\Gamma(s)}\cr
&-4 {\Gamma(s-{1 \over 2}) \over \Gamma(s)}
 \sum_{r=0}^\infty {{{1 \over 2}-s} \choose r}
{m^{-2r-2s+1} \over (2 \pi)^{2r+2}} (2r+1)! (1-2^{-2r-1}) \zeta(2r+2)
\bigr]\cr}.$$
After restoring the appropriate
factors of $\rho$, this simplifies to the following expression:
\beq {W_c(\rho) \over N}= -{1 \over 24 \rho^3}
-{1 \over 4 \rho^3} \sum_{r=0}^\infty {{1 \over 2} \choose r}
{1 \over \pi^{2r+2}} (2r+1)! (1-2^{-2r-1}) \zeta(2r+2).\eeq
(In the above equation and in \eqn{bseries}, $\zeta(2r+2)$ is the
Riemann zeta function.)
\vfill
\eject
\n {\bf $O(N)$ sigma model at finite temperature}

The examples considered above in this section described quantum phase
transitions at zero temperature.
The extensions to finite temperatures can be easily worked out. We will
just state the results here, for the cases
$S^1 \times S^1 \times S^1$ (a torus, which has zero curvature) and
$S^2 \times S^1$ (constant positive curvature).
In the example $S_\rho^1 \times S_\rho^1 \times S_\rho^1$, we find that
$b=0$ and
the critical value of $m$ is a solution to the equation,
\beq
\int\limits_0^\infty dt \quad t^{-{3 \over 2}} e^{-m^2t}
[\theta_3^3(0, e^{-{\rho^2 \over 4t}})-1]+m \Gamma(-{1 \over 2}) =0\eeq
where $\Gamma(-{1 \over 2})=-2 \sqrt \pi$ and the Theta function,
$\theta_3(0,q)$, has the power series representation,
$\theta_3(0,q)=\sum\limits_n q^{n^2}, |q|<1$.
We can compute the free energy using this critical value of $m$.
In the case where the manifold is $M=S^2 \times S^1$,
at the critical point, we find that $b=0$. The critical value of $m$ is
a solution to the following (finite)
equation and the free energy can be evaluated with this value of $m$.
\beq
\sum_{n=-\infty}^\infty {\rm P} \int\limits_{-\infty}^\infty dx
({x \over 2 \rho} {\rm cosec}{x \over 2 \rho}-1)
{K_1 ({\sqrt{m (x^2+n^2 \beta^2)}}) \over {\sqrt{ (x^2+n^2 \beta^2)}}}
- {4 \pi \over m\beta} {\rm Log}2\sinh({m \beta \over 2})=0\eeq
where $K_1(x)$ is the MacDonald function.

\n {\bf Discussion}~~ The following table
is a summary of results obtained on various
manifolds at the critical point.
$m$ denotes the mass,  $b$ the spontaneous magnetization and $W$ is the
regularized free energy density.
{\vskip0.1in
\def\entry#1:#2:#3:#4:#5:{\strut\quad#1\quad&\quad#2\quad&\quad#3\quad&
\quad#4\quad&\quad#5\quad\cr}
\offinterlineskip\tabskip=0pt \halign{%
\vrule\quad\hfill#\hfil\quad\vrule&\quad\hfill#\hfil\quad\vrule&
\quad\hfill#\hfil\quad\vrule&\quad\hfill#\hfil\quad\vrule&
\quad\hfill#\hfil\quad\vrule\cr
\noalign{\hrule}
\vphantom{\vrule height 2pt}&&&&\cr\noalign{\hrule}
\entry \it Manifold:\it $m$:\it $\langle \phi(x) \phi(y) \rangle$:
\it $b$:\it $W$:
\vphantom{\vrule height 2pt}&&&&\cr\noalign{\hrule}
\vphantom{\vrule height 2pt}&&&&\cr\noalign{\hrule}
\entry $R^3$: zero: power law: zero: zero:
\entry $R^2\times S_{\beta}^1$: $2\log\tau \over \beta$:
exponential decay: zero:
$\not= 0$:
\entry $S^1\times S^1\times R$: $\not= 0$: exponential decay: zero: $\not= 0$:
\entry $S^2\times R$: zero: exponential decay: zero: zero:
\entry $H_{\rho}^2\times R$: ${1/2\rho}$: power law in  $u$:
$\not= 0$: $\not= 0$:
\entry : : (exponential  in $d$): : :
\vphantom{\vrule height 2pt}&&&&\cr\noalign{\hrule}}}

These results, obtained by explicit calculations, show
some surprises. The first line in the table
is well--known; it is just the statement that the $O(N)$ $\sigma$-model has a
second order phase transition. The spontaneous magnetization vanishes and the
correlation length diverges at the transition point.  In fact the `mass' (
vacuum expectation value of the $\sigma$ field)
is in that case the inverse of the
correlation length. But in more general curved geometries, this relation need
not hold. For example in the case $S^2\times R$, the correlation length is
actually finite although $m=0$.
 It is finite in the $S^2$ direction because of the finite size
of $S^2$. In the $R$  direction, the correlation decays exponentially.
This case is conformally equivalent to $R^3$ which explains why $m,b$ and the
free energy density vanish at criticality.
On the
other hand for the manifold $H^2\times R$, the parameter $m\neq 0$;  in fact
the value of $m$ is such that
correlation decays like an exponential (in the geodesic distance)  in all three
 directions.
The cases $R^2\times S^1$ and
$S^1\times S^1\times R$ have zero curvature and seem to fit with naive
expectations. In the case $H^2\times R$ we
also
find that at the critical point the spontaneous magnetization is non--zero.
This could be relevant to the  phase transition in two dimensional anti--
ferromagnets (such as copper oxides\[cuo])  when the lattice is subject to an
 external stress that bends it into a hyperboloid.

\vskip0.1in
\n {\bf Appendix A: Conformal Geometry in Three Dimensions}
\vskip0.1in
Let $(M,g)$ be an oriented  Riemannian manifold of dimension
$d$. We are mostly interested in metrics of positive
(Euclidean) signature. Two
Riemann metrics $g,\tilde g$  are said to be conformally
equivalent if there is a  smooth function $f:M\to R$ such that
$\tilde g=e^{2f} g$. Under such a transformation of the
metric, $g\to e^{2f}g$, the angles between two vectors will be
unchanged, but the lengths of vectors will change. A metric is
conformally flat if it is conformally equivalent to a flat
metric.The set of conformal transformations form an
abelian group, the group of smooth functions $C^{\infty}(M)$
under addition.

 We can define a  `conformal structure' on an
oriented manifold $M$ to be a non--degenerate symmetric tensor
density $\hat g_{ij}$ of weight $-2$ and with $\det \hat g=1$.
(We find it convenient to
use the convention that the volume is a density of weight $d$).
 The determinant of $\hat g_{ij}$ is a scalar, so the
condition that it be equal to one is diffeomorphism invariant.
 Given a Riemannian metric, a conformal structure is
determined by $\hat g_{ij}=g^{-{1\over d}} g_{ij}$, where $g$
is
the determinant of $g_{ij}$. Clearly $\hat g_{ij}$ is
invariant if we change $g_{ij}\to e^{2f}g_{ij}$.  $\hat
g_{ij}$ determines the equivalence class of the metric tensor
under conformal transformations.  It is reasonable to call
$\hat g_{ij}$ the metric tensor density, since it determines
the angle $\theta(u,v)$  between two vectors:
\beq
     \cos\theta(u,v)={\hat g_{ij}u^iv^j\over \sqrt[\hat
g^{kl}u^ku^l \hat g^{mn}v^mv^n]}.
\eeq
This is a scalar although the `length' of a vector  $\surd
[\hat
g_{kl}u^ku^l]$ is a scalar density of weight $-2$.

 If ${\hat \Gamma}$ is the space of conformal structures on
$M$,
the group $\diff M$ acts on it by pull--back:
\beq
     \phi\in \diff M, \phi:{\hat \Gamma}\to {\hat \Gamma},
\hat g\mapsto
\tilde g=\phi^*\hat g.
\eeq
Explicitly in terms of co--ordinates,
\beq
     \tilde g_{ij}(x)=[\det \pdr \phi]^{-{2\over
d}}{\hat g}_{kl}(\phi(x)){\pdr \phi^k\over \pdr
x^{i}}{\pdr \phi^l\over \pdr x^{j}}.
\eeq
Two metric densities  that differ only by such a
diffeomorphism are to
be regarded as equivalent, as they only differ by `change of
co--ordinates'. We will thus be interested in objects of
conformal geometry that are covariant under this action of the
group  of diffeomorphisms $\diff M$ of $M$. These objects will
be defined on the space $\hat \Gamma/\diff M$ of equivalence
classes of conformal structures under  the diffeomorphism
group. This space however, is not in general a manifold, since
the action of the group can have  fixed points.

It is useful to study first  the   change of $\hat g_{ij}$
under the
action of an
infinitesimal
 diffeomorphism. This  is the Lie derivative with respect to
a vector
field,
\beq
     [\Lie_v \hat g]_{ij}= v^k\pdr_k\hat g_{ij}+\pdr_iv^k\hat
g_{kj}+\pdr_{j}v^k\hat
g_{ik}-{2\over d}\pdr_kv^k\hat g_{ij}.
\eeq
The last  term arises from the fact that $\hat g_{ij}$ is a tensor {\it
density}, of weight $-2$.

By the way, this might also be thought of as a covariant
derivative that
maps the covariant vector density $\hat v_i=\hat g_{ij}v^j$ to
a symmetric traceless tensor density. After some calculation,
\beq
     [\Lie_v \hat g]_{ij}=(D\hat v)_{ij}=\pdr_i\hat
v_j+\pdr_j\hat
v_i-{2\over
d}\pdr_k\hat v_l\hat g^{kl}\hat g_{ij}-2 \hat
\Gamma^{k}_{ij}\hat
v_k
\eeq
Here,
\beq
     \hat \Gamma^k_{ij}=\half \hat g^{kl}[\pdr_i\hat
g_{lj}+\pdr_j \hat g_{il}-{2\over d}\hat g^{mn} \hat g_{ij}
\pdr_{m}\hat g_{nl}].
\eeq
Our  construction shows that  $D$ is a covariant derivative
that maps covariant vector densities of weight $-2$ to
symmetric
traceless covariant tensor  densities of weight $-2$.  $\hat
\Gamma^{k}_{ij}$ are the conformal geometric analogues of the
Christoffel symbols of Riemannian geometry. It would be interesting
to understand conformal curvature
as a `commutator' of such conformally covariant derivatives.

If $[\Lie_v \hat g]_{ij}=0$, $v^i$ is a conformal Killing
vector;
then this infinitesimal  diffeomorphism
has $\hat g_{ij}$ as a fixed point.
The compact manifold with the maximum number of
such conformal Killing vectors
is $S^n$; they form the Lie algebra  $\un{O(n+1,1)}$.
This space can also be thought of as $R^n$;
the conformal killing vectors correspond then to
translations, rotations, dilatations and
some `special conformal transformations':
\beq
	v_i=a_i+\theta_{ij}x^j+\lambda x_i+x^2b_i-2x_ib.x.
\eeq

The quotient space $\hat \Gamma/\diff M$ will not then be a
Hausdorff topological space and hence not a manifold. This
difficulty can be avoided by
restricting to the open dense subset of $\hat \Gamma$ which
does not have any conformal Killing vectors. Alternatively, we
can restrict to the subgroup $\diff_p M$ which agrees with
the identity map up two derivatives, at one point $p$.
Even in the case with largest number of conformal Killing vectors, this
condition
will remove all of them.
This will remove all conformal Killing vectors when $M$ is
compact. There could still be fixed points due to  finite conformal
isometries; but they only lead to `orbifold' type
singularities in the quotient which can be removed by passing
to a covering space. All objects of interest in conformal
geometry will be sections of vector bundles over the space
$\Q=\hat \Gamma/\diff_p M$. A similar point of view was found
to be very useful in studying Yang--Mills
theories
\[ym2+theo].

There is another point of view on conformal structures in low
dimensions that is useful.
Recall that on any oriented manifold the Levi--Civita symbols
$\eps_{i_1\cdots i_d}$ and $\eps^{i_1\cdots i_d}$ are natural
anti--symmetric   tensor densities  of weight  $-d$ and $d$
respectively. If $d=2$, this allows us to describe a conformal
structure also by a tensor $J_a^b=\hat g_{ac}\eps^{cb}$. We
can see now that $J_a^bJ_b^c=\det \hat g \eps_{ae}\eps^{ec}=-
\delta_a^c$. This shows that a conformal structure is the same
as a complex structure in dimension two (the integrability
condition on $J_a^b$ is trivial in two dimensions).
The analogue when $d=3$ is a tensor density $c_i^{jk}=\hat
g_{il}\eps^{ljk}$ of weight $1$ satisfying
\beq
c_i^{jk}c_m^{il}+{\rm cyclic}[jkl]=0,\quad \det [c_i^{jk}c_j^{il}]=1
.\eeq
This  will define a `cross product'  among
covariant vector densities of weight $-1$.
The conformal metric can be recovered from the
contraction $\hat g^{ij}=-\half c_l^{ki}c_{k}^{lj}$.
Thus in three
dimensions a conformal structure is the same as a Lie algebra
structure, isomorphic to
$\un{SU}(2)$, (or $\un{SU(1,1)}$) on vector densities
of weight $-1$. This point of view needs to be investigated further.

Returning to the study of $\Q$, let us first get an idea of
how big $\Q$ is. We have a principal bundle $\diff_p M\to \hat
\Gamma\to \Q$.  Let us ask what conditions a one--form in
$\hat \Gamma$ must satisfy in order that it be the pullback of
a one--form in $\Q$ by the natural projection $\pi:\hat
\Gamma\to \Q$. This will give an idea of the size of the
cotangent space of $\Q$.
A tangent vector to $\hat \Gamma$ is a traceless symmetric
covariant tensor density of weight $-2$ in $M$.  A cotangent
vector (one-form) in $\hat \Gamma$ is a traceless symmetric
contravariant tensor density of   weight $d+2$. The
contraction of a vector $h$ in $\hat\Gamma$ with a one--form
$t$ is then,
\beq
     i_h t=\int_M t^{ij}h_{ij} d^d x.
\eeq
Given a vector field in $M$, there is a vertical vector field
in $\hat \Gamma$, given by the infinitesimal action:
\beq
     h_{ij}=[D\hat v]_{ij}
\eeq
where $D$ is the covariant derivative of $\hat v_i=\hat
g_{ij}v^j$ defined earlier. If a  one--form $t$ in $\hat
\Gamma$ is the pull--back of a one--form in $\Q$ it must
annihilate all vertical vector fields:
\beq
     \int_M t^{ij}[Dv]_{ij}d^dx =0
\eeq
This implies that the covariant divergence of $t^{ij}$ should
be zero:
\beq
     [D^*t]^i=0
\eeq \label{div}
(Strictly speaking, $[D^*t]^i$ must be zero  when contracted
with vector fields that vanish up two derivatives at $p$.
This means that $[D^*t]^i$ is
a combination of derivatives of the  delta function
concentrated at $p$. If the manifold does not admit conformal
Killing vectors, this subtlety can be
ignored.)

This covariant divergence $[D^*t]^i$ can be defined in terms
of $\hat g_{ij}$ alone:
\beq
     [D^*t]^i=\pdr_it^{ij}-\hat \Gamma^j_{ik}t^{ik}
\eeq
$\hat \Gamma^j_{ik}$ being the conformal analogues of the
Christoffel symbols defined earlier.
The covariant divergence of such a tensor density has a
meaning within conformal geometry, without any reference to Riemann metric.
This  divergence is in
fact a vector density of weight $d+2$.

So far we have considered $t_{ij}$ as a one--form at a point
$\hat g_{ij}$ of $\hat \Gamma$. If it is the pull--back of a
one--form in $\Q$, $t_{ij}$ must change along the vertical
direction in a way that is determined by the action of
$\diff_p M$:
\beq
     [\Lie_v t]^{ij}+\int {\delta t^{ij}\over \delta \hat
g_{kl}}(y)[D\hat v]_{kl}(y) d^d y=0.
\eeq \label{vert}
Conversely, any  one--form in $\hat\Gamma$ satisfying
the above conditions is the pullback of a one--form in
$\Q$.

The size of $\Q$ is given by the number of independent
solutions to the equation, ~$[D^*t]^i=0$, among traceless
symmetric tensor densities. A traceless tensor density has
{}~${(d+2)(d-1)\over 2}$ independent components at each point of
$M$. The condition of having zero divergence is given by $d$ equations
at each point of $M$. Hence the number of independent
components in a one--form of $\Q$ is ${(d-2)(d+1)\over 2}$.
A way to formalize this statement is that $\Q$ is
a manifold modelled over a vector space
$[C^{\infty}(M)]^{(d-2)(d+1)\over 2}$.

The conditions of being traceless and divergence free
are familar properties of   stress tensor of a conformal field
theory. Indeed the stress tensor of a
conformal field theory is a one-form on $\Q$,
when there is no `conformal anomaly'(for example in
three dimensions). Moreover, in general, the stress tensor will
be a closed 1--form; it is not always exact. If the parity anomaly
vanishes it is exact.
Similar considerations arise in the canonical formalism of
general relativity \[mtw].

 In
particular, if $d=2$,there are no local degrees of freedom in
such a $t^{ij}$: the equations $[D^*t]^i=0$ form an elliptic
system. This does not mean that $\Q$ is trivial, just that it
is finite dimensional. The dimension of $\Q$ (the number of
independent solutions of $[D^*t]^i=0$ ) is $6h-6$ for a
compact manifold of genus $h\geq 2$.
The fact that $\Q$ has no local degrees of freedom does imply
that all 2--manifolds are locally conformally equivalent and
hence  conformally
flat.

In the case of most interest to us, $d=3$, there are two
independent degrees of freedom per point of $M$. Thus there
must be a conformal curvature
tensor, which measures the local deviation from conformal
flatness. We will discuss this tensor soon.

\n {\bf Conformal Geometry from Riemannian Geometry}

We can view the space $\hat\Gamma$ as the quotient of the space of
Riemannian metrics $\Gamma$
by the group $C^{\infty}(M)$.
The principal bundle $C^{\infty}(M)\to \hat \Gamma\to \Gamma$
is defined by the group action
\beq
	g_{ij}\to e^{2f}g_{ij}.
\eeq
Thus it is
also possible to study conformal geometry by looking at
structures in Riemannian geometry  invariant under   the
 semi--direct product $\diff M\semi C^{\infty}(M)$.

 The curvature (or Riemann) tensor transforms as follows
under a conformal transformation\[eisenhart]:
\beqs{
     R_{ijkl}\to \tilde R_{ijkl}&=e^{2f}[R_{ijkl}+
g_{il}f_{jk}+g_{jk}f_{il}-g_{ik}f_{jl}-g_{jl}f_{ik}\cr
     &+(g_{il}g_{jk}-g_{ik}g_{jl})|df|^2
.}\eeqs
Here,
\beq
f_{ij}=\nabla_i\pdr_jf-\pdr_if\pdr_jf,\quad
|df|^2=g^{mn}f_{,m}f_{,n}.
\eeq
(For explicit computations there is still nothing better than
the classical co--ordinate notation of tensor calculus.)
{}From this, it follows that the Ricci tensor and Ricci scalar
transform as follows:
\beq
     R_{ij}\to \quad \tilde R_{ij}= R_{ij}+(d-
2)f_{ij}+g_{ij}[\Delta f+(d-2)|df|^2].
\eeq
\beq
     R\to \tilde R=e^{2f}[R+2(d-1)\Delta f+(d-1)(d-2)|df|^2]
\eeq
Here, $\Delta f=g^{ij}f_{,ij}$ is the Laplacian.
The traceless part of the Riemann  tensor,
\beq
     C^h_{ijk}=R^h_{ijk}+{1\over d-2}[\delta^h_jR_{ik}-
\delta^h_k R_{ij}+g_{ik}R^h_j-g_{ij}R^h_k ]+{1\over (d-1)(d-
2)}[\delta^h_k g_{ij}-\delta^h_jg_{ik}]R
\eeq
is called the Weyl tensor. It is identically zero unless
$d\geq 4$. The Weyl tensor is invariant  under
conformal transformations:
\beq
     C^h_{ijk}\to C^h_{ijk}.
\eeq
If $d\geq 4$, a manifold is conformally flat if and
only if
the Weyl tensor vanishes.

If $d=1$ every manifold is flat, so in particular conformally
flat. If $d=2$, every manifold is locally conformally
equivalent to flat space.
This was already seen in the previous discussion on $\Q$.

 When $d=3$, the situation is more subtle: the Weyl tensor
vanishes, but not every 3--manifold is even locally conformally
flat. This is already clear from the previous discussion of
the number of degrees of freedom of $\Q$. There is a tensor
density, special to three dimensions that
measures conformal curvature\[eisenhart]:
\beq
     \Omega^{ij}=g^{1\over 3}\eps^{ikl}\nabla_k[R^j_l-{1\over
4}\delta^j_lR]
\eeq
Here $\eps^{ijk}$ is the Levi--Civita tensor (which depends
on  the  choice of orientation on $M$) and $g=\det g_{ij}$.
This is a tensor density  of weight $5$ that is invariant
under conformal
transformations.

Thus $\Omega^{ij}$ is a geometric object  on the space $\hat
\Gamma=\Gamma/C^{\infty}(M)$: it is in fact a one--form.
Moreover, a 3--manifold is locally conformally flat iff
$\Omega^{ij}=0$. In fact it is now  possible to verify that
it satisfies the conditions
\beq
     \Omega^{ij}=\Omega^{ji},\quad \hat g_{ij}\Omega^{ij}=0,\quad
[ D^*\Omega]^{i}=0.
\eeq
Moreover, it satisfies the condition that it is a  geometrical
tensor, depending only on $g_{ij}$,
\beq
     \Lie_{v} \Omega^{ij}+\int {\delta \Omega^{ij}\over \delta
g_{kl}(y)}\Lie_{v}g_{kl}(y)d^3y=0
.\eeq
 These conditions have a simple interpretation: $\Omega^{ij}$
defines a 1--form on $\Q$.
In fact it is a closed form. This can be verified
by explicit computation of its exterior derivative.
A better strategy is to use the fact \[deseretal+wit]
that the conformal curvature is the derivative of the Chern--Simons term.

Thus there is something special about conformal geometry in
three dimensions: the conformal curvature  is given by  a
tensor density peculiar to three dimensions. Also, the space
of conformal structures carries a 1--form that measures the
deviation of each point from local conformal flatness. There
are two independent components in $\Omega^{ij}$ per point of $M$.

$\Omega^{ij}$ being zero only implies local conformal flatness.
Typically, for compact $M$, there is a finite dimensional
manifold of inequivalent conformal structures  with $\Omega^{ij}=0$
on some manifold $M$. This finite dimensional space is
analogous to Teichmuller space. The equations $\Omega^{ij}=0$ are
the conformal analogues of the field equations of  Chern--
Simons   theory; the
solutions are parameterized by conjugacy classes  of
homomorphisms of the fundamental group of $M$ to the conformal
isometry group of $R^3$, which is $O(4,1)$  (recall that the
Teichmuller  space of a Riemann surface surface is the set of
conjugacy classes of homomorphisms $\pi_1(\Sigma)\to O(2,1)$)
However, the study of the theory on such conformally flat
manifolds does not seem to give sufficiently detailed
information on phase transition.
The major difference between two and three dimensional
conformal field theory is that in three dimension, the effect
of conformal curvature needs to be taken into account.
This is why we believe that generalizations of conformal field
theory to three (or higher) dimensions
based on invariance under $O(n+1,1)$ describe only part of the story.

\n {\bf Appendix B: Poisson sum formula on $S^2$}
\vskip0.1in
The Poisson sum formula \eqn{poisson1} is quite standard. We give here
a brief derivation of the Poisson sum formula \eqn{poisson2}:
We start with the general Poisson sum formula,
\beq {1 \over \beta} \sum_n e^{-{4 \pi^2 t \over \beta^2} (n+\sigma)^2}
e^{2 \pi i{x \over \beta}(n+\sigma)}
={1 \over {\sqrt {4 \pi t}}} \sum_n e^{-{(x+n \beta)^2 \over 4 t}}
e^{-2 \pi i n \sigma}\eeq
If we choose $\sigma={1 \over 2}$ and $\beta=2 \pi$, and differentiate
 both sides of the above equation w.r.t $x$, we obtain,
\beq {i \over 2 \pi} \sum_n (n+{1 \over 2})
e^{-t(n+{1 \over 2})^2} e^{i x (n+{1 \over 2})}
=-{1 \over 2 t {\sqrt {4 \pi t}}} \sum_n (x+2 \pi n)
e^{-{(x+ 2 \pi n)^2 \over 4 t}} e^{-\pi i n }\eeq \label{a0}
\n Recall that,
$$\sum_{l={1 \over 2}{\rm integers}} {\rm sgn}(l) e^{ilx}
=\sum_{l={1 \over 2}}^\infty (e^{ilx}- e^{-ilx})=i {\rm cosec}{x \over 2}$$
Therefore,
\beq {\rm sgn}(l)
={i\over 2 \pi} \int\limits_0^{2 \pi} dx \quad {\rm cosec}{x \over 2}
e^{-ilx}\eeq \label{a3}
\n Use equation \eqn{a3} to write \eqn{a0} as,
$${1 \over 2 \pi} \sum_{l={1 \over 2} {\rm integers}} l {\rm sgn}(-l)
e^{-l^2 t}=-{1 \over (4 \pi t)^{3 \over 2}} \sum_{n=-\infty}^\infty (-1)^n
\int\limits_0^{2 \pi} (x+2 \pi n) dx
{\rm cosec}{x \over 2} e^{-{(x+2 \pi n)^2 \over 4 t}}$$
On letting $x+2 \pi n \rightarrow x$ and on further simplification, we get,
\beq {1 \over 2 \pi} \sum_{l={1 \over 2}}^\infty l e^{-l^2 t}
={1 \over (4 \pi t)^{3 \over 2}} \sum_{n=-\infty}^\infty
\int\limits_{2 \pi n}^{2 \pi n + 2 \pi} dx \quad
{x \over 2} {\rm cosec}{x \over 2} e^{-{x^2 \over 4t}}\eeq
Rescaling $t$, $t \rightarrow {t \over \rho^2}$, and extracting completely
the small $t$ divergence of the integral,
$${1 \over 2 \pi} \sum_{l={1 \over 2}}^\infty l e^{-{l^2 \over \rho^2} t}
={\rho^2 \over (4 \pi t)^{3 \over 2}}
\int\limits_{-\infty}^\infty dx \quad ({x \over 2 \rho}
{\rm cosec}{x \over 2 \rho}-1) e^{-{x^2 \over 4 t}} + {\rho^2 \over 4 \pi t}$$
which is the required Poisson sum formula \eqn{poisson2}.

\n {\bf Acknowledgements} We thank ~R.~J.~Henderson, ~F.~Lizzi,
{}~S.~Sachdev,  ~G.~Sparano and ~O.~T.~Turgut for discussions.
P.V. would like to thank the Dept. of Physics at the Univ. of Rochester
for hospitality and
S.G. thanks the organizers of the
conference MRST-'94 (Montreal) for the opportunity to present part of
this work.
This work was supported in part by the US Dept. of Energy, Grant No.
DE-FG02-91ER40685.

\vskip0.1in
\n {\bf REFERENCES}
\vskip0.1in
\n \[arefeva] I.Ya.Aref'eva, Teor. Mat. Fiz. 31(1977) 3;
I.Ya. Aref'eva and S.I.Azakov, Nucl. Phys. B162(1980) 298

\n \[polyabook] A.M.Polyakov, Gauge fields and strings, (Harwood Academic
Publishers, 1987)

\n \[sachdev] S.Sachdev, Phys. Lett. B309(1993) 285

\n \[rosenstein] B.Rosenstein, B.J.Warr, S.H.Park, Nucl. Phys. B116(1990) 435

\n \[zinnjustin] J.Zinn-Justin, Quantum field theory and critical phenomena,
(Oxford Univ. Press, 1989)

\n \[jain] S.Jain, Int. J. Mod. Phys. A3 (1988) 1759

\n \[witten1] E. Witten, {\it The Central Charge in Three Dimensions}, in
{\it Physics and Mathematics of Strings}, L. Brink, D. Friedan
and A. M. Polyakov Eds., World Scientific, Singapore (1990)

\n \[osborn] H. Osborn, Phys. Lett. B222 (1989) 97

\n \[latorre] A.Cappelli, D.Friedan and J.I.Latorre, Nucl.Phys. B352(1991) 616;
A. Cappelli, J. I. Latorre and X. Vilasis-Cardona,
Nucl. Phys. B376 (1992) 510

\n \[mottola] I. Antoniadis, P. Mazur and E. Mottola, Nucl. Phys. B388(1992)
627

\n \[ferretti] G.Ferretti and S.G.Rajeev, Phys. Rev. Lett. 69 (1992) 2033

\n \[birrell] N.D.Birrell and P.C.W.Davies, Quantum fields in curved space,
(Cambridge Univ. Press, Cambridge, 1982)


\n \[davies] E.B.Davies, Heat kernels and spectral theory, (Cambridge Univ.
Press, Cambridge, 1989)

\n \[parker] T.Parker and S.Rosenberg, J.Differential Geometry 25(1987) 199

\n \[fradkin] A.H.Castro Neto and E.Fradkin, Nucl. Phys. B400(1993) 525

\n \[cardy] J.L.Cardy, Nucl. Phys. B290(1987) 355; Nucl. Phys. B366(1991) 403

\n \[lang] S.Lang, $SL_2(R)$, (Springer-Verlag New York co., 1985)

\n \[cuo] A.V.Chubukov, S.Sachdev and J.Ye, Preprint COND-MAT/9304046 and
related references therein.

\n \[ym2+theo] S.G.Rajeev, Phys. Lett. B212 (1988) 203;
K.S.Gupta, R.J.Henderson, S.G.Rajeev and O.T.Turgut,
J. Math. Phys. 35 (1994) 3845

\n \[mtw] C.W.Misner, K.S.Thorne, J.A.Wheeler, Gravitation, (W.H.Freeman and
Company, 1973)

\n \[eisenhart] L.P.Eisenhart, Riemannian Geometry, (Princeton Univ. Press,
1949)

\n \[deseretal+wit] S. Deser, R. Jackiw and S. Templeton, Ann. Phys. 140 (1982)
372; J.H.Horne and E.Witten, Phys. Rev. Lett. 62 (1989) 501

\end